\documentclass[prb,preprint]{revtex4-1} 

\usepackage{amsmath}  
\usepackage{amsfonts} 
\usepackage{graphicx} 
\usepackage{braket}
\usepackage[colorinlistoftodos]{todonotes}
\usepackage{comment}
\usepackage{verbatim}

\begin{document}

\title{A Non-Relativistic Model of Heavy Quarkonium}

\author{Christian Farina}
\email{cfarina@illinois.edu} 
\affiliation{Department of Physics, University of Illinois, Urbana-Champaign, IL 61801}


\date{\today}

\begin{abstract}
    We develop a non-relativistic, spin-dependent, phenomenological model of heavy quarkonium. The model is based on a hydrogen-like Schr\"odinger equation derived from a spin-dependent Hamiltonian that includes a hyperfine spin--spin interaction term. The resulting system is solved using matrix mechanics by diagonalizing the Hamiltonian to obtain the energy spectrum. The energies are further optimized using variational methods. Spin--orbit corrections are included at first order in perturbation theory. The resulting spectrum shows good agreement with experimental data for conventional heavy quarkonium states. The framework combines matrix mechanics, variational methods, and perturbation theory, and is therefore suitable for students familiar with quantum mechanics.
\end{abstract}
\maketitle

\section{Introduction and Motivation}
\label{sec:Intro}
One of the most well-studied problems in both elementary and advanced treatments of quantum mechanics is that of a particle in a central potential. The resulting time-independent, three-dimensional Schr\"odinger equation is typically solved in spherical coordinates, reducing it to a one-dimensional radial differential equation:
\begin{equation}
-\frac{\hbar^2}{2\mu}\left[\frac{d^2}{dr^2}+\frac{2}{r}\frac{d}{dr} - \frac{\ell(\ell+1)}{r^2}\right]R(r) + V(r)R(r) = ER(r),
\label{eq:CentralHamiltonian}
\end{equation}
where $\mu = \frac{m_e m_p}{m_e + m_p}$ is the reduced mass of the two-particle system. The most prominent example is the hydrogen atom, for which $V(r) = -\frac{e^2}{4\pi \epsilon_0 r}$ (see, for example, \cite{Griffiths,Townsend}).

The hydrogen atom is a particularly important case, not only because of its physical relevance, but also because it is one of the few systems that admits an exact analytical solution; the other notable example being the simple harmonic oscillator. In general, however, even relatively simple systems such as the helium atom require approximation and numerical methods. One such approach is the variational method (see \cite{Sakurai}, p.~313), in which one selects a trial state $\ket{\psi(\beta)}$ depending on a parameter $\beta$, and computes the expectation value of the Hamiltonian $\braket{\psi|H|\psi}$ to obtain an upper-bound estimate for the ground state energy, $E_0$. The energy is then minimized by varying with respect to the parameter:
\begin{equation}
   \frac{\partial}{\partial \beta}\frac{\braket{\psi|H|\psi}}{\braket{\psi|\psi}}=0.
   \label{eq:varParams}
\end{equation}
Another approach is to use matrix methods, in which the wavefunction is expanded in a finite (generally non-orthonormal) basis $\{\ket{\psi_n}\}$, truncated at some $N_{\text{max}}$:
\begin{equation}
\ket{\psi} = \sum_{n=1}^{N_{\text{max}}} c_n \ket{\psi_n}.
\end{equation}
Substituting into the Schr\"odinger equation and projecting onto $\bra{\psi_m}$ leads to the generalized eigenvalue problem
\begin{equation}
\sum_n H_{mn} c_n = E \sum_n S_{mn} c_n,
\label{eq:GenEigen}
\end{equation}
where
\begin{equation}
H_{mn} = \bra{\psi_m} H \ket{\psi_n}, \qquad S_{mn} = \braket{\psi_m|\psi_n}.
\end{equation}
The matrix $S_{mn}$ is the overlap matrix, which reduces to the 
identity only if the basis is orthonormal. Solving this generalized 
eigenvalue problem yields $N_{\max}$ approximate energy eigenvalues 
and their corresponding eigenvectors. For a given set of quantum 
numbers $S$, $L$, and $J$, the eigenvalues are arranged in increasing 
order, with the lowest eigenvalue corresponding to the lowest-energy 
state with those quantum numbers and the higher eigenvalues 
corresponding to successive radial excitations. The eigenvalues may 
therefore be identified with the $n=1,2,\ldots$ radial states for a 
given $S$, $L$, and $J$. In this way, the full spectrum is constructed 
by solving the generalized eigenvalue problem separately for the 
different combinations of $S$, $L$, and $J$.

In practice, the basis expansion is truncated at a finite value of 
$N_{\mathrm{max}}$, since the variational energies rapidly converge 
and additional basis functions eventually lead to increasing linear 
dependence and diminishing numerical improvement.

These methods are particularly useful when the potential $V(r)$ is analytically intractable and have proven effective even in more complex few-body systems \cite{Farina:2020slb}. Matrix-based approaches have also been applied previously to heavy quarkonium systems in pedagogical contexts \cite{Belhouari2023}, demonstrating that numerical diagonalization methods provide an accessible framework for studying bound-state spectroscopy. For small corrections to known solutions, such as relativistic effects and spin-dependent terms, perturbation theory provides a complementary and widely used approach.

In this paper, we combine matrix methods, variational methods and time-independent, first-order perturbation theory, to compute the full mass spectrum of heavy quarkonium (which is explained below). Throughout the paper, we use natural units, in which $\hbar=1$ and $c=1$ unless otherwise stated. The paper is organized as follows: In Section \ref{sec:QCD} we describe the basic properties of heavy quarkonium, quantum chromodynamics and quark models. The model chosen is presented in Section \ref{sec:Model}, along with the reasons for the radial basis choice. Section \ref{sec:Results} contains the theoretical results, which are compared to the experimental values of all well-established quarkonium states. Section \ref{sec:Summary} contains a summary and conclusions along with some suggested improvements that could be implemented by students.

\section{Heavy Quarkonium, QCD and Constituent Quark Models}
\label{sec:QCD}
The term quarkonium refers to a bound state of a quark and an antiquark which combine to form a hydrogen-like state. Quarks are the fundamental components of all matter and they interact through the strong force to form composite particles called hadrons. There are six different quarks called \emph{up}, \emph{down}, \emph{strange}, \emph{charm}, \emph{bottom}, and \emph{top}, in increasing order of mass (usually labeled by \emph{u, d, s, c, b, t}). Besides carrying electric charge, quarks possess an additional quantum number known as color charge, and are the only elementary particles that participate in the strong interaction. The labeling by quark type is referred to as \emph{flavor}. Quarks of different flavor behave identically under the strong interaction, but differ in their masses and in how they participate in the electroweak interaction. Because these mass differences are large, nature does not treat all flavors equally, so flavor symmetry is only approximate in nature.

Hadrons are separated into baryons (states made of three quarks) and mesons (states made of a quark and an anti-quark). The theory that describes the properties of quarks is called quantum chromodynamics (QCD for short). In the theory of QCD, the strong interaction between quarks is mediated by massless force carriers called gluons. This is analogous to the interaction of charged particles in quantum electrodynamics (QED - the quantum theory of electromagnetic phenomena), in which the force between electrons is mediated by the massless photons. 

QCD is much more difficult to work with than QED primarily due to two of its key features: its non-Abelian gauge structure and the running of its coupling constant, $\alpha_s$, which determines the strength of the interaction and varies with energy. The non-Abelian nature of the theory implies that gluons, in addition to mediating the strong force between quarks, also interact with one another. This self-interaction enables the existence of exotic hadronic states, such as glueballs, composed entirely of gluonic degrees of freedom, and hybrid mesons, which contain both valence quarks and gluonic excitations \cite{MeyerSwanson2015}. These states are very relevant to current research in non-perturbative QCD and hadron spectroscopy as they could shed light on the role of gluonic degrees of freedom at low energy, but they are also much more difficult to describe and will not be treated here.

The energy dependence of the coupling also has important physical consequences. At high energies (short distances), the coupling becomes small, leading to a regime in which quarks behave approximately as free, point-like particles. This phenomenon is known as \emph{asymptotic freedom}. At low energies (large distances), however, the coupling grows rapidly, resulting in a confining interaction that prevents quarks from being isolated. This property is known as \emph{confinement}, and our understanding of it remains largely non-perturbative and phenomenological. Due to these challenges, one must often resort to effective modeling when describing bound states of quarks.

One of the most successful approaches to hadron phenomenology is the \emph{constituent quark model} (CQM) \cite{PDG2026,Entem:2025bqt}, in which hadrons are described in terms of their minimal valence quark content---three quarks for baryons such as protons and neutrons, and a quark--antiquark pair for mesons. Constituent quark models, particularly non-relativistic ones, have a long history \cite{GodfreyIsgur1985,IsgurKarl1978,IsgurKarl1979} and, despite their differences, share several common features: (i) the inclusion of an effective potential that mimics confinement, and (ii) spin-dependent interactions that account for energy splittings among states with different spin and angular momentum.

Heavy quarkonium states, composed of either charm or bottom quark--antiquark pairs (commonly known as charmonium and bottomonium), are particularly well-suited for non-relativistic analyses due to the large masses of the constituent quarks. And since they are bound states of two particles with equal mass, a non-relativistic description leads to a Schr\"odinger equation that is very similar to that of hydrogen, but at a reduced length scale. Moreover, the experimentally observed spectra of charmonium and bottomonium have been studied extensively within the potential model framework, and theoretical predictions are found to be in good agreement with experimental data.

\section{Model Development}
\label{sec:Model}
The starting point of our model is the heavy-quark Hamiltonian 
\begin{equation}
    H=H_0+V(r),
\end{equation}
where 
$H_0$ is the kinetic part, given by 
\begin{equation}
    H_0=2m_q-\frac{1}{m_q}\Big(\frac{d^2}{dr^2}+\frac{2}{r}\frac{d}{dr}\Big)-\frac{\ell(\ell+1)}{m_q r^2},
\end{equation}
and the potential contains spin-independent and spin-dependent terms, $V(r)=V_{SI}(r)+V_{SD}(r)$, with
\begin{align}
    V_{SI}&=-\frac{4}{3}\frac{\alpha_{s}}{r}+br+\mathcal{C}, \nonumber \\
    V_{SD}&=\frac{32\pi \alpha_{s}}{9m^2_q}\tilde{\delta}_{\sigma}(r)\mathbf{S_q\cdot S_{\bar{q}}}+\frac{1}{m^2_q}\Big( \frac{2 \alpha_{s}}{3r^3}-\frac{b}{2r} \Big)\mathbf{S\cdot L}.
\end{align}
This non-relativistic model is justified by the large mass of the constituent quarks, which suppresses relativistic corrections. The spin-independent part of the potential contains the standard Coulomb-plus-linear form, describing one-gluon exchange at short distances and confinement at large distances, typical of quark models. The use of a Coulomb-plus-linear (Cornell) potential is strongly motivated by lattice QCD calculations of the static quark--antiquark potential, which exhibit a short-distance Coulombic behavior and a linearly rising confining interaction at large separations \cite{Bali2001,JugeKutiMorningstar1999}. The inclusion of the $2m_q$ term ensures that the eigenvalues correspond to the total meson masses rather than binding energies. The $\alpha_s$ is a Coulomb parameter analogous to the electromagnetic structure constant $\alpha_{e}$ in electrodynamics, $b$ is the confining parameter also known as the string constant, and $\mathcal{C}$ is a constant shift.  

The spin-dependent part includes a spin-spin hyperfine interaction, modeled by a Gaussian-smeared contact term, and a spin–orbit interaction. These terms can be formally derived from a Breit–Fermi reduction \cite{EichtenFeinberg1981} of the one-gluon-exchange interaction. The reduction also generates a tensor interaction term of the form $\frac{1}{m^2_q}\frac{4 a_{s}}{r^3}\left( 3\mathbf{\left(S_q\cdot \hat{r}\right)\left( S_{\Bar{q}}\cdot \hat{r} \right)}-\mathbf{S_q\cdot S_{\Bar{q}}} \right)$, whose matrix elements are non-vanishing only for $L>0$ spin-triplet states. In this work, the tensor contribution is neglected, as it is expected to be small for the states considered, and also because students at this level might be less familiar with tensor operators. The smeared delta function is taken to be
\begin{equation}
\tilde{\delta}_{\sigma}(r)=\left(\frac{\sigma}{\sqrt{\pi}}\right)^3 e^{-\sigma^2 r^2},
\end{equation}
which regularizes the short-distance behavior and accounts for the finite spatial extent of the constituent quarks. Since $\sigma$ has units of inverse distance, $1/\sigma$ is a measure of the range of the interaction. Quarks and antiquarks have intrinsic spin $1/2$, hence they combine to form mesons of total spin $S=0$ or $S=1$. This can be used to rewrite the hyperfine term by re-expressing the spin dot product in terms of the total spins 
\begin{equation}
    \mathbf{S_q\cdot S_{\bar{q}}}=\frac{\mathbf{S_{tot}^2}-\mathbf{S_{q}^2}-\mathbf{S^2_{\bar{q}}}}{2}
\end{equation}
and using $\mathbf{S^2_j}\ket{\psi}=S_j(S_j+1)\ket{\psi}$, with $S_j=0,1$ for the total spin and $S_j=1/2$ for the individual spins. This leads to 
\begin{equation}
    \mathbf{S_q\cdot S_{\bar{q}}}\ket{\psi}=\frac{1}{2}\left(S(S+1)-\frac{3}{2}\right)\ket{\psi} = 
    \begin{cases} 
    -\frac{3}{4}\ket{\psi}, \quad S = 0 \\ +\frac{1}{4}\ket{\psi}, \quad S = 1, 
    \end{cases}
\end{equation}
which shows that the hyperfine term is attractive for $S=0$ and repulsive for $S=1$ mesons. The repulsive nature of the potential pushes the quarks apart and increases the total energy of $S=1$ mesons, making them heavier than their $S=0$ counterparts, for which the potential is attractive. 

The spin–orbit term differs from the standard Thomas precession form due to the assumed scalar nature of the confining potential, leading to a relative sign difference. This form is motivated by early quark model studies by Kang and Schnitzer \cite{KangSchnitzer1975}, which demonstrated that a scalar confining interaction can reproduce the observed pattern of spin--orbit splittings in heavy mesons. At short distances, the $1/r^3$ behavior introduces a divergence, which is regulated by introducing a cutoff $r_0=0.1~\text{GeV}^{-1}$ , replacing $1/r^3$ with $1/r_0^3$ for $r < r_0$. The lower limit $r_0$ was chosen by decreasing it until the calculated spin-dependent corrections became insensitive to further reductions. Thus, $r_0$ serves only as a short-distance numerical regulator that prevents the model from probing a region where the effective Hamiltonian should not be trusted and is not treated as
an additional fitted parameter. Following the same procedure outlined above, it can be shown that the $\mathbf{S\cdot L}$ operator has matrix elements equal to $\braket{\mathbf{S\cdot L}}=\frac{1}{2}\left[J(J+1)-S(S+1)-L(L+1) \right]$, which are non-zero only for $S=1$, $L\geq1$.

Lastly, since the spin–orbit interaction is subleading compared to the hyperfine term, it is treated in first-order perturbation theory by diagonalizing the system and calculating the $\braket{\psi|V_{SL}|\psi}$ corrections numerically (for arbitrarily small values of $r_0$ using any numerical integration method such as Simpson rule or a quadrature rule) which are then added to the total energy.

\begin{figure}[h]
\centering
\includegraphics[]{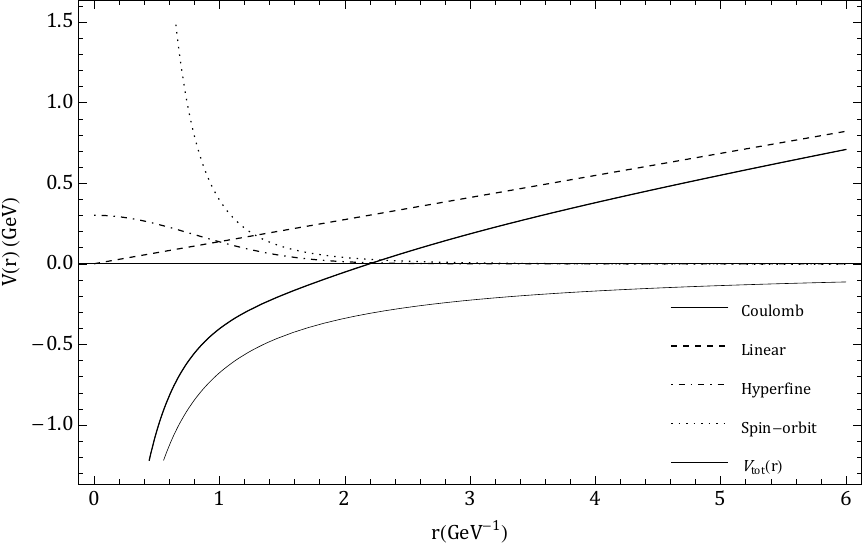}
\caption{Plot of the total potential and the various individual terms. Only the $r$-dependent part of each potential is plotted, as both the hyperfine and spin-orbit have different factors depending on the eigenvalues of the angular momentum operators. The total potential (solid line) is $V_{tot}(r)=-\frac{4}{3}\frac{a_{s}}{r}+br+\frac{32\pi a_{s}}{9m^2_q}\left(\frac{\sigma}{\sqrt{\pi}}\right)^3 e^{-\sigma^2 r^2}$, while the spin-orbit term $\frac{1}{m^2_q}\Big( \frac{2 a_{s}}{3r^3}-\frac{b}{2r} \Big)$ is displayed as a thin dotted line. The values of the parameters are those listed in Table \ref{tab:params}.}
\label{fig:totPotential}
\end{figure}

The radial dependence of the total potential for charmonium is shown in Figure \ref{fig:totPotential}, where the individual terms are also shown separately, highlighting the short-range nature of the hyperfine part. The total potential includes all of the terms that were used for the diagonalization (Coulomb plus linear plus hyperfine) but does not include the spin-orbit term which was treated perturbatively. The latter is separately shown in the plot as a thin dotted line. 
\subsection{Parameters Optimization}
The optimized parameters were determined by fitting the model to the experimentally established quarkonium spectrum through a chi-squared minimization procedure. The charmonium and bottomonium sectors were fitted independently. To reduce the influence of spin-dependent splittings during the initial stage of the fit, we first fixed the spin-dependent smearing parameter $\sigma$ and the constant mass shift $\mathcal{C}$, and determined the remaining parameters using spin-averaged masses.

The spin-averaged masses were constructed by taking weighted averages over states with the same orbital angular momentum $L$, where the weights correspond to the spin degeneracy factor $(2J+1)$. For example, the spin-averaged mass for the $S$-wave $(L=0)$ states is
\begin{equation}
\overline{M}_{S}=
\frac{
1\cdot M({}^1S_0)+3\cdot M({}^3S_1)
}{4}.
\end{equation}
For the $P-$wave, the full spin-averaged mass including the ${}^1P_1$ states is
\begin{equation}
\overline{M}_{P}=
\frac{
1\cdot M({}^3P_0)
+3\cdot M({}^3P_1)
+5\cdot M({}^3P_2)
+3\cdot M({}^1P_1)
}{12}.
\end{equation}
This is the most general way of averaging all the $P-$wave states. Alternatively, one could just consider the ${}^3P_1$ states and use instead
\begin{equation}
\overline{M}_{P}=
\frac{
1\cdot M({}^3P_0)
+3\cdot M({}^3P_1)
+5\cdot M({}^3P_2)
}{9}.
\label{eq:Pcentroid}
\end{equation}
In this paper, we use the second choice because it is the standard choice in heavy quarkonium models. The reason for this choice is that the ${}^1P_1$ state $h_c$ lies already very close to the weighted average of Eq. \ref{eq:Pcentroid}. Similarly, the spin-averaged masses for the higher-angular-momentum multiplets were constructed using the appropriate degeneracy-weighted averages. Once the quark parameters $m_q$, $\alpha_S$ and $b$ were determined, they were fixed and the spin-averaged spectrum was again fit to determine the mass shift, $\mathcal{C}$. Since both $m_q$ and $\mathcal{C}$ directly change the value of the total mass of the meson and both shift the overall spectrum, fitting them separately avoided parameter correlations and degeneracies.

Once the spin-independent parameters were obtained, the spin-dependent parameter, $\sigma$, was adjusted to reproduce the observed hyperfine and fine-structure splittings by fitting to the full spectrum. The values used in this computation are given below and are consistent with values commonly used in constituent quark models.
\begin{table}[h]
    \centering
    \begin{tabular}{c|ccccc}
    \hline\hline
      Quarkonium & ~~$m_q$ (GeV)~~ & ~~$\alpha_s$~~ & ~~$b~\text{(GeV)}^2$~~ & ~~$\sigma$ (GeV)~~ & ~~$\mathcal{C}$ (GeV)~~ \\
      \hline
      Charmonium & 1.545 & 0.508 & 0.137 & 0.897 & -0.113 \\
      Bottomonium & 4.793 & 0.390 & 0.167 & 1.638 & 0.0 \\ 
      \hline\hline
    \end{tabular}
    \caption{Values for the parameters used in the model. The values are optimized by fitting to the experimentally available charmonium and bottomonium spectra.}
    \label{tab:params}
\end{table}

\subsection{Radial Basis Choice and Energies Computation}
At this point, the only matter left to settle is the choice of basis in which to diagonalize and minimize the resulting radial system. There are many possible choices and all come with their own advantages and pitfalls. We employ a Gaussian Expansion Method (GEM) \cite{Hiyama:2003cu}, in which the radial wavefunctions are expanded in a non-orthogonal Gaussian basis with a discrete set of Gaussian width parameters:
\begin{equation}
\psi(r)=\sum_{n=1}^{N_{\mathrm{max}}}c_n\psi_n(r,\beta,\ell)
\end{equation}
where
\begin{equation}
\psi_n(r,\beta,\ell)=\mathcal{N}_{\ell,\beta,n}r^{\ell}e^{-n\beta^2r^2/2}
\end{equation}
In the above, $\beta$ is a variational parameter, $N_{\max}$ is the number of terms in the expansion, $n=1,\ldots,N_{\max}$ labels the individual basis functions, and $\ell$ is the orbital angular momentum of the meson, and $\mathcal{N}_{\ell,\beta,n}$ the normalization equal to
\begin{equation}
    \mathcal{N}_{\ell,\beta,n}=\sqrt{\frac{2 n^{\ell+\frac{3}{2}}\beta ^{2 \ell+3}}{\Gamma\left( \ell+\frac{3}{2}\right)}},
\end{equation}
where $\Gamma(n)$ is the Gamma function. The factor $r^\ell$ ensures the correct near-origin behavior of the radial wavefunctions for nonzero orbital angular momentum. This choice is in line with what has been historically done for conventional meson and baryon models, in which authors have used Gaussian expansions as an alternative to harmonic-oscillator (HO) bases and other orthogonal polynomial bases and more sophisticated hyperspherical methods \cite{Richard:1992uk,Arifi:2024mff}.

\begin{figure}[h]
\centering
\includegraphics[width=\textwidth]{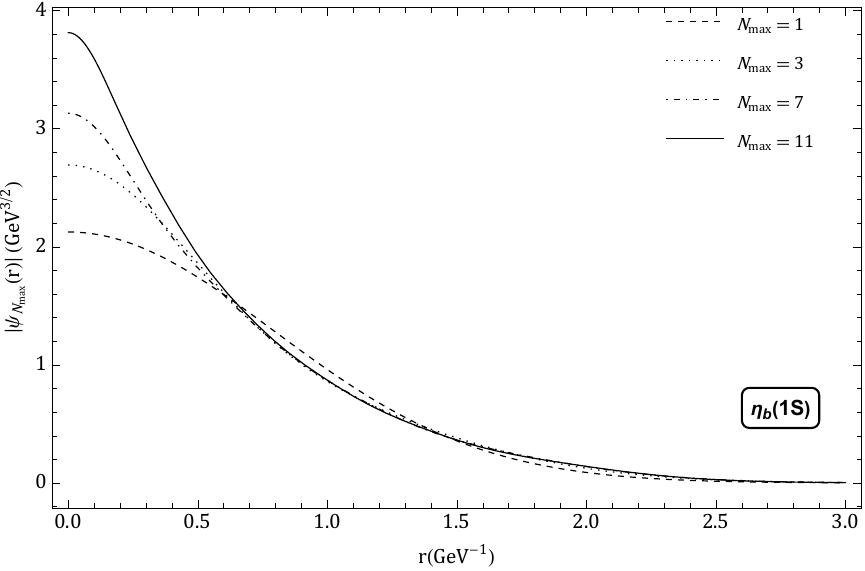}\\
\includegraphics[width=\textwidth]{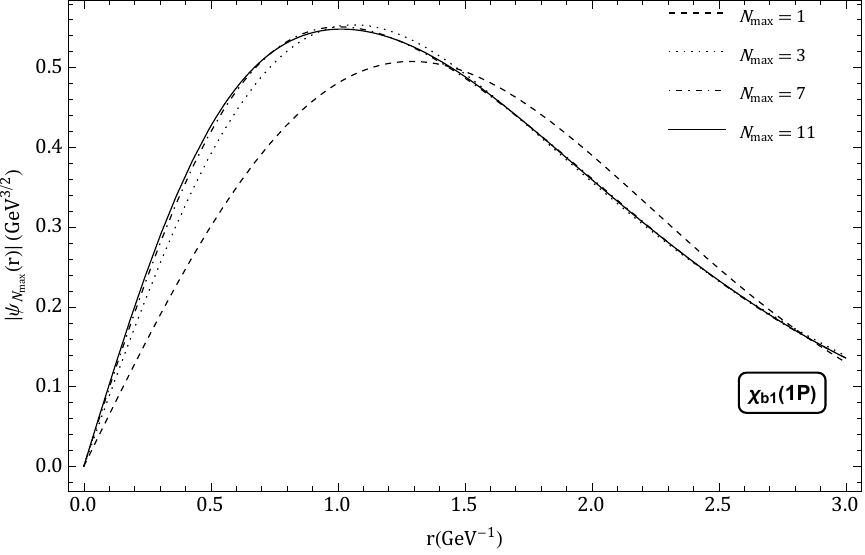}
\caption{
Convergence of the $\eta_b(1S)$ and $\chi_{b1}(1P)$ radial
wavefunctions as the number of Gaussian basis functions is increased. For a detailed discussion see the text.
}
\label{fig:wfconvergence}
\end{figure}

The matrix elements then give integral expressions of the form
\begin{equation}
H_{mn}=\braket{\psi_m|H|\psi_n}=\int_0^\infty dr r^2 \psi^*_{m}(r;\beta,\ell)H(r)\psi_{n}(r;\beta,\ell).
\label{eq:Hmatrixelements}
\end{equation}
The main advantage of this basis is that it yields analytic or semi-analytic evaluation of most matrix elements in the Hamiltonian, with the remaining terms evaluated numerically. Moreover, the convergence is sufficiently rapid that only a relatively small number of basis functions ($N_{\mathrm{max}}= 11$ in our case) are required to obtain accurate energy estimates. This rapid convergence arises because linear combinations of Gaussian functions provide an efficient representation of smooth bound-state wavefunctions over a wide range of distances. As additional basis functions are included, the largest changes occur primarily near the origin, where the short-range Coulombic and hyperfine interactions are strongest, while the asymptotic tail stabilizes more quickly due to the coherent superposition of the functions. Consequently, the basis saturates rapidly and the overall functional form of the wavefunction changes only slightly as $N_{\mathrm{max}}$ increases. This behavior is shown graphically for the $\eta_{b}(1S)$ and $\chi_{b1}(1P)$ states in Figure \ref{fig:wfconvergence}.

However, the figure also shows that the $S-$wave does not converge rapidly enough in the $r<0.6~\text{GeV}^{-1}$ regime, where Coulombic and hyperfine effects are strongest. While not ideal, this is not unexpected: the variational principle ensures that energies generally converge faster than wavefunctions. Additionally, Gaussian functions have difficulties reproducing $S-$wave Coulombic cusps at the origin due to their vanishing derivative. On the contrary, the $P-$wave wavefunctions, which are less susceptible to hyperfine effects and more sensitive to spin-orbit ones, show excellent overall convergence as $N_{\max}$ is increased. 

For each of the two states in Fig. \ref{fig:wfconvergence} we have computed the dominant short-range matrix elements as functions of $N_{\max}$, reported in Table \ref{tab:sd-convergence}. For the $\eta_{b}(1S)$, the hyperfine matrix element changes by approximately $4.5\%$ between $N_{\max}=7$ and $N_{\max}=11$ and by $1.6\%$ between $N_{\max}=10$ and $N_{\max}=11$. While the convergence is slower than desired, it only corresponds to a few MeV in the hyperfine contribution to the mass. On the other hand, for the $\chi_{b1}(1P)$, the hyperfine matrix element decreases substantially with increasing $N_{\max}$, but its absolute magnitude is already negligible. More importantly, the spin-orbit matrix element, which provides the relevant short-range correction for this state, is considerably more stable, changing by only $0.25\%$ between $N_{\max}=10$ and $N_{\max}=11$.

\begin{table}[h]
    \centering
    \begin{tabular}{c|ccc}
        \hline\hline
        $N_{\max}$ 
        & $\langle V_{\mathrm{hyp}}\rangle_{\eta_b(1S)}$ (MeV)
        & $\langle V_{\mathrm{hyp}}\rangle_{\chi_{b1}(1P)}$ (MeV)
        & $\langle V_{LS}\rangle_{\chi_{b1}(1P)}$ (MeV) \\
        \hline
        7  & 78.8478 & 0.54665 & 13.9333 \\
        8  & 79.3845 & 0.43038 & 14.0376 \\
        9  & 80.6074 & 0.31449 & 14.0780 \\
        10 & 81.1171 & 0.25099 & 14.1301 \\
        11 & 82.4066 & 0.19542 & 14.1650 \\
        \hline\hline
    \end{tabular}
    \caption{Convergence of the radial hyperfine and spin-orbit matrix
    elements with the number of Gaussian basis functions $N_{\max}$
    for the $\eta_b(1S)$ and $\chi_{b1}(1P)$ states.}
    \label{tab:sd-convergence}
\end{table}
The main drawback of the GEM is that the expansion functions are not 
orthogonal, leading to a generalized eigenvalue problem of the type 
described above in Eq. \ref{eq:GenEigen}. In practice, this comes with no additional conceptual complexity, since the system is diagonalized using standard linear algebra routines (we implemented the diagonalization with the Eigen/Dense package in C++). 

The minimization of the resulting energies is carried out with respect to the variational parameter, $\beta$, according to 
Eq. \ref{eq:varParams}. For each value of $N_{\max}$, $\beta$ is varied independently and the value that minimizes the energy of the state under consideration is selected. Consequently, the optimal value of $\beta$ generally changes as $N_{\max}$ is increased. The values obtained for the $\eta_c(1S)$ and $\eta_b(1S)$ states are reported explicitly in Table \ref{tab:Econvergence}. Explicit expressions for the matrix elements are given in Section \ref{sec:appendix}. Here, we show the full, step-by-step derivation of the confining term, $br$, for pedagogical purposes. The matrix element is given by:
\begin{equation}
    \braket{\psi_m|br|\psi_n}=\int_0^\infty dr r^2 \psi^*_{m}(r;\beta,\ell)(br)\psi_{n}(r;\beta,\ell).
\end{equation}
Plugging in the explicit expression for the wavefunctions yields
\begin{equation}
    \braket{\psi_m|br|\psi_n}=\frac{2b\beta^{2\ell+3}\sqrt{(mn)^{\frac{3}{2}+\ell}}}{\Gamma\left( \ell+\frac{3}{2}\right)}\int_0^\infty dr r^{3+2\ell}e^{-\frac{\beta^2r^2(m+n)}{2}}. 
\end{equation}
Here, $\ell$, $m$ and $n$ are integers, therefore the argument of the integral can be recast into the form $r^k e^{-\beta^2r^2}$, with $k=2\ell+3$ and $\beta^2=(m+n)\beta^2/2$. The final expression can then be obtained using the well-known result
\begin{equation}
    \int_0^\infty r^k e^{-\beta^2r^2}=\frac{1}{2\beta^{k+1}}\Gamma{\left( \frac{k+1}{2}\right)}.
\end{equation}
Re-expressing everything in terms of $\ell$, $m$ and $n$ and using $k+1=2(\ell+2)$ and $\Gamma{\left( \frac{k+1}{2}\right)}=\Gamma{\left(\ell+2\right)}$ yields 
\begin{equation}
\int_0^\infty dr\,r^{3+2\ell}
e^{-\frac{\beta^2r^2(m+n)}{2}}
=
\frac{1}{2}
\left(\frac{2}{(m+n)\beta^2}\right)^{\ell+2}
\Gamma(\ell+2).
\end{equation}
Lastly, substituting back into the original expression, canceling the factor of 2 and combining the powers of $\beta$ gives
\begin{equation}
    \braket{\psi_m|br|\psi_n}=\frac{2^{\ell+2} b (m n)^{\frac{(2 \ell+3)}{4}}\Gamma (\ell+2)}
   {\beta  \left( m+n\right)^{\ell+2}\Gamma \left(\ell+\frac{3}{2}\right)}.
\end{equation}
If desired, the expression can be further simplified by re-expressing the Gamma functions in terms of factorials and double factorials, but we find this form to be more useful for computation. All the other terms in the Hamiltonian can be computed in a similar way. While the intermediate algebra may appear tedious, the full calculation can be efficiently automated in a symbolic computation package such as Wolfram Mathematica, which we used to compute the matrix elements.
\section{Results and Comparison to Experimental Values}
\label{sec:Results}
In this section we present the results of the model and compare them to the experimental spectrum of charmonium and bottomonium mesons. The masses are listed in Tables \ref{tab:charmonium_comparison} and \ref{tab:bottomonium_comparison}, along with the percent error. The experimental values are taken from the Particle Data Group \cite{PDG2026}. The overall agreement is very good, with most percent errors remaining below $1\%$.

The bottomonium spectrum is reproduced with smaller deviations than charmonium. This reflects the increased validity of the nonrelativistic approximation in the heavy-quark limit. Because the bottom quark is significantly heavier than the charm quark, the typical quark velocities in bottomonium are smaller, and relativistic corrections are correspondingly reduced. In addition, bottomonium probes shorter distance scales where the strong coupling is weaker, so the interaction is more strongly influenced by the short-distance Coulombic term in the potential.

Charmonium, by contrast, lies in an intermediate regime where confinement effects, relativistic corrections, and other nonperturbative contributions play a more significant role. These effects are not fully captured within the present potential model, leading to somewhat larger discrepancies. The improved agreement in bottomonium therefore serves as a useful consistency check of the model and illustrates the domain of validity of the nonrelativistic approach.
We have omitted several states whose interpretation as conventional $q\bar{q}$ mesons is not firmly established, and for which mixing, threshold effects, and possible multiquark structure become important. Examples include the $\psi(4230)$ and $Z_c(4430)$, which have been discussed in the literature as candidates for tetraquark, hybrid, or hadronic molecular states (see, for instance, \cite{WangZhao2025}). Including such states would require going beyond the simple quark–antiquark potential framework used here. 

\begin{figure}[h]
\centering
\includegraphics[]{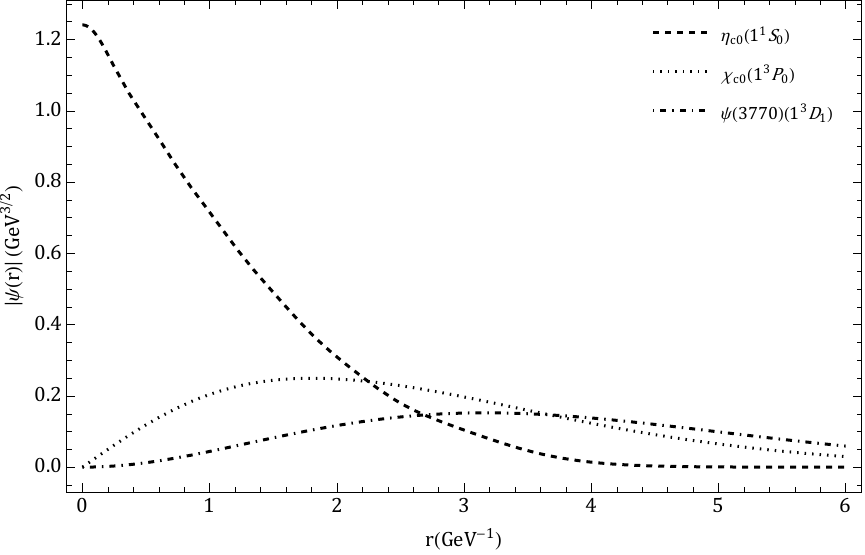}
\caption{Plot of $|\psi(r)|$ vs $r$ for three states with different orbital angular momentum.}
\label{fig:wavefuncts}
\end{figure}

Analyzing the $r-$dependence of $\psi(r)$ is also crucial, because the differing behavior of the wavefunctions near the origin plays a significant role in determining the relative importance of the spin-dependent interactions.
Within the charmonium sector, the states with the largest errors are the $\chi_{c0}(1P)$, the $\psi(4360)$, and the $\psi_3(3842)$, and in general, $P$ and $D$-wave states tend to have larger errors. The short-range spin–spin (hyperfine) interaction is dominant near the origin and therefore has its largest effect on $S$-wave states ($L=0$), whose wavefunctions are nonzero at $r=0$. This contribution is suppressed for $P$-wave states ($L=1$), such as the $\chi_{c0}(1P)$, whose wavefunctions vanish at the origin. As a result, the fine-structure splittings in these states are governed primarily by spin–orbit and tensor interactions. The same reasoning applies to $D$-wave states such as the $\psi_3(3842)$ ($S=1$, $L=2$, $J=3$). This is shown in Figure \ref{fig:wavefuncts}, where the absolute values of the wavefunctions are plotted as functions of $r$. 

The identification of the $\psi(4360)$ is less certain, with assignments corresponding to either the $n=3$ or $n=4$ radial excitation appearing in the literature \cite{Ding:2007rg}, which is the reason for the question mark used in the table.

\begin{table}[h]
\centering
\begin{tabular}{ccccc}
$~{n}^{2S+1}L_{J}~$ & ~Name(Multiplet)~ & ~Experimental Mass~ & Theoretical Mass & ~$|\%~ error|$~\\
\hline\hline
${1}^1 S_0$ & $\eta_c(1S)$ & 2.9839 & 2.9720 & 0.40 \\
${1}^3 S_1$ & $J/\psi(1S)$ & 3.0969 & 3.1129 & 0.52 \\
\hline
${1}^3 P_0$ & $\chi_{c0}(1P)$ & 3.4147 & 3.5293 & 3.36 \\
${1}^3 P_1$ & $\chi_{c1}(1P)$ & 3.5107 & 3.5270 & 0.47 \\
${1}^1 P_1$ & $h_c(1P)$ & 3.5254 & 3.5197 & 0.16 \\
${1}^3 P_2$ & $\chi_{c2}(1P)$ & 3.5562 & 3.5225 & 0.95 \\
\hline
${2}^1 S_0$ & $\eta_c(2S)$ & 3.6375 & 3.6505 & 0.36 \\
${2}^3 S_1$ & $\psi(2S)$ & 3.6861 & 3.6624 & 0.64 \\
\hline
${1}^3 D_1$ & $\psi(3770)$ & 3.7737 & 3.7960 & 0.59 \\
${1}^3 D_2$ & $\psi_2(3823)$ & 3.8237 & 3.7912 & 0.85 \\
${1}^3 D_3$ & $\psi_3(3842)$ & 3.8427 & 3.7840 & 1.53 \\
\hline
${2}^3 P_0$ & $X(3915)$ & 3.9217 & 3.9232 & 0.04 \\
${2}^3 P_2$ & $\chi_{c2}(2P)$ & 3.9223 & 3.9253 & 0.08 \\
\hline
${3}^3 S_1$ & $\psi(4040)$ & 4.0390 & 4.0472 & 0.20 \\
${2}^3 D_1$ & $\psi(4160)$ & 4.1910 & 4.1389 & 1.24 \\
\hline
${3}^3 P_1$ & $\chi_{c1}(4274)$ & 4.2740 & 4.2550 & 0.44 \\
${3}^3 S_1$ & $\psi(4360)$ & 4.3680 & 4.4351 & 1.54 \\
${4}^3 S_1$ & $\psi(4415)$ & 4.4210 & 4.3687 & 1.18 \\
${5}^3 S_1$ & $\psi(4660)$ & 4.6300 & 4.6540 & 0.52 \\
\hline\hline
\end{tabular}
\caption{Comparison of experimental and theoretical charmonium masses (in GeV) using optimized parameters. The average deviation is approximately 26.8 MeV.}
\label{tab:charmonium_comparison}
\end{table}

\begin{table}[h]
\centering
\begin{tabular}{ccccc}
$~{n}^{2S+1}L_{J}~$ & ~Name(Multiplet)~ & ~Experimental Mass~ & Theoretical Mass & ~$|\%~ error|$~\\
\hline\hline
${1}^1 S_0$ & $\eta_b(1S)$ & 9.3987 & 9.3847 & 0.15 \\
${1}^3 S_1$ & $\Upsilon(1S)$ & 9.4603 & 9.4508 & 0.10 \\
\hline
${1}^3 P_0$ & $\chi_{b0}(1P)$ & 9.8594 & 9.9107 & 0.52 \\
${1}^3 P_1$ & $\chi_{b1}(1P)$ & 9.8928 & 9.9104 & 0.18 \\
${1}^1 P_1$ & $h_b(1P)$ & 9.8993 & 9.9085 & 0.09 \\
${1}^3 P_2$ & $\chi_{b2}(1P)$ & 9.9122 & 9.9097 & 0.03 \\
\hline
${2}^1 S_0$ & $\eta_b(2S)$ & 9.9990 & 9.9992 & $<0.01$ \\
${2}^3 S_1$ & $\Upsilon(2S)$ & 10.0233 & 10.0033 & 0.20 \\
\hline
${1}^3 D_1$ & $\Upsilon(1D)$ & 10.1611 & 10.1503 & 0.11 \\
${1}^3 D_2$ & $\Upsilon_2(1D)$ & 10.1637 & 10.1495 & 0.14 \\
\hline
${2}^3 P_0$ & $\chi_{b0}(2P)$ & 10.2325 & 10.2502 & 0.17 \\
${2}^3 P_1$ & $\chi_{b1}(2P)$ & 10.2555 & 10.2503 & 0.05 \\
${2}^1 P_1$ & $h_b(2P)$ & 10.2598 & 10.2500 & 0.10 \\
${2}^3 P_2$ & $\chi_{b2}(2P)$ & 10.2686 & 10.2504 & 0.18 \\
\hline
${3}^3 S_1$ & $\Upsilon(3S)$ & 10.3552 & 10.3350 & 0.19 \\
\hline
${4}^3 S_1$ & $\Upsilon(4S)$ & 10.5794 & 10.6023 & 0.22 \\
\hline\hline
\end{tabular}
\caption{Comparison of experimental and theoretical bottomonium masses (in GeV) using optimized variational parameters. The average deviation is approximately 14.6 MeV.}
\label{tab:bottomonium_comparison}
\end{table}
Table \ref{tab:Econvergence} shows the convergence of $M$ as a function of the number of terms in the expansion, $N_{\max}$, for the $S$-wave ground states $\eta_c(1S)$ and $\eta_b(1S)$, respectively. In both cases, the expansion saturates at $N=11$, but this form makes it hard to determine whether the basis exhibits good convergence toward the experimental values. 

\begin{figure}[h]
\centering
\includegraphics[width=\linewidth]{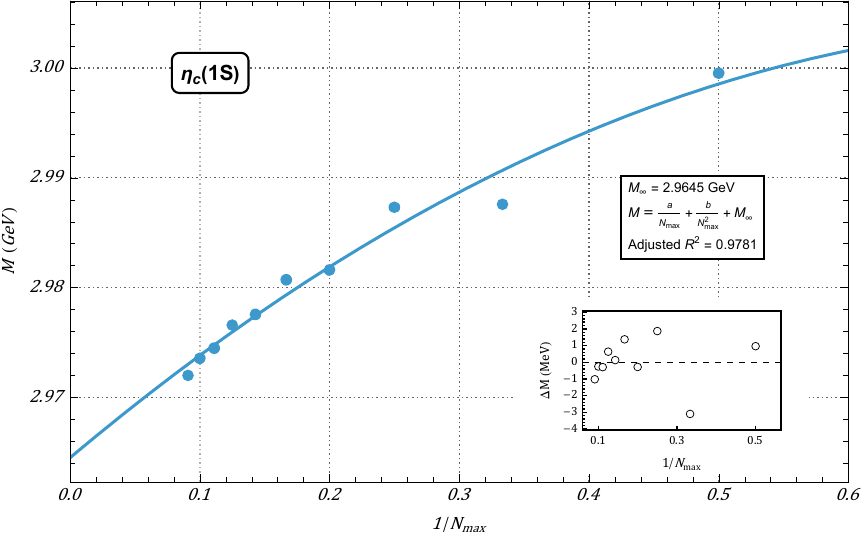}\\
\includegraphics[width=\linewidth]{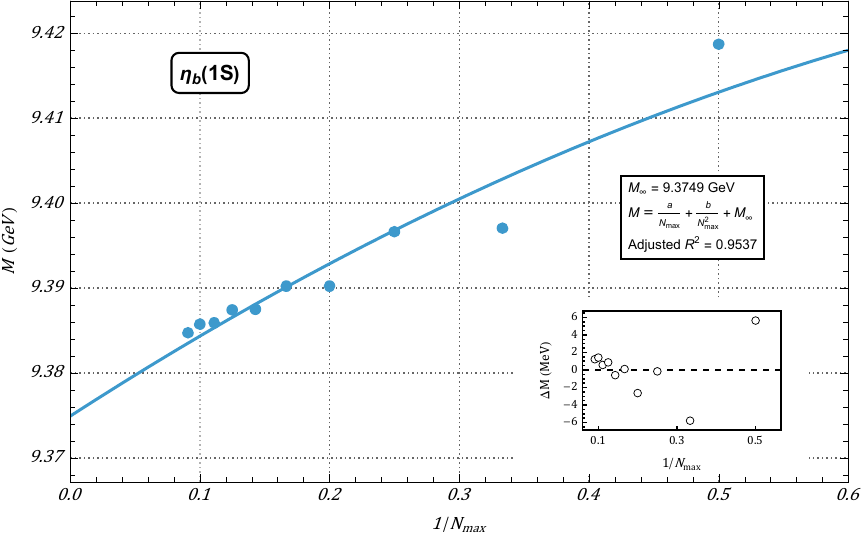}
\caption{Convergence of the theoretical mass as a function of $1/N_{\max}$, for the $\eta_{c}(1S)$ (above) and the $\eta_{b}(1S)$ (below). The blue fitted line gives the asymptotic value for the mass, $M_{\infty}$, as the number of terms approaches infinity. The values of the parameters in each case are $a=0.0994$, $b=-0.0627$ for the $\eta_{c}(1S)$ and $a=0.0988$, $b=-0.0488$ for the $\eta_{b}(1S)$.}
\label{fig:Econvergence}
\end{figure}

Empirically, the variational energies obtained with a Gaussian basis often exhibit approximately inverse-power convergence with the number of basis functions. We therefore fit the calculated masses to the equation 
\begin{equation}
    M(N_{\max})=M_{\infty}+\frac{a}{N_{\max}}+\frac{b}{N^2_{\max}},
\end{equation}
where $M_{\infty}$ represents the estimated infinite-basis limit, shown in Fig.\ref{fig:Econvergence}. The quality of the fit is assessed using the adjusted $R^2$ and computing the residuals. The choice of this functional form is empirical rather than derived analytically; it was selected because it provides, in both cases, an excellent description of the observed convergence and yields small residuals with no systematic dependence on $1/N_{\max}$.

The systematic tendency for the converged $S$-wave masses to lie slightly below the experimental values may indicate that the effective Coulomb strength parameter $\alpha_s$ is somewhat overestimated. Since the Coulomb interaction dominates at short distances, its effects are strongest for $S$-wave states, whose wavefunctions remain nonzero at the origin. However, since this behavior is also present for some of the higher angular momentum states, in both charmonium and bottomonium, an alternative explanation for the lower theoretical masses could be found in the limitations of the model itself, as discussed below.

\begin{table}[h]
    \centering
    \begin{tabular}{c|cc||cc}
    \hline\hline
   & \multicolumn{2}{c||}{$\eta_c(1S)$} & \multicolumn{2}{c}{$\eta_b(1S)$} \\
     ~$N_{\max}$~ & ~$M$ (GeV) & $\beta$ (GeV)~ & ~$M$ (GeV) & $\beta$ (GeV)~ \\
    \hline\hline
        1 & 3.0012 & 0.710 & 9.4281 & 1.260 \\
        2 & 2.9995 & 0.675 & 9.4187 & 1.130 \\
        3 & 2.9879 & 0.725 & 9.3971 & 1.195 \\
        4 & 2.9873 & 0.735 & 9.3966 & 1.170 \\
        5 & 2.9816 & 0.760 & 9.3902 & 1.215 \\
        6 & 2.9807 & 0.775 & 9.3901 & 1.220 \\
        7 & 2.9776 & 0.795 & 9.3875 & 1.250 \\
        8 & 2.9766 & 0.810 & 9.3874 & 1.260 \\
        9 & 2.9745 & 0.825 & 9.3859 & 1.285 \\
        10 & 2.9736 & 0.835 & 9.3857 & 1.295 \\
        11 & 2.9720 & 0.845 & 9.3847 & 1.320 \\
        \hline
        $\rightarrow\infty$ & 2.9650 & & 9.3749 &  \\
    \hline\hline
    \end{tabular}
    \caption{Convergence of the theoretical mass as a function of the number of terms in the expansion, $N$, for the $\eta_c(1S)$ and $\eta_b(1S)$. The variational parameters obtained are shown in the second and fourth columns. The last row is obtained from the fits shown in Fig. \ref{fig:Econvergence}}
    \label{tab:Econvergence}
\end{table}
The variational energies exhibit stable convergence as the number of basis functions increases. For fixed model parameters, the extrapolated value $M_{\infty}$ may be interpreted as an estimate of the infinite-basis eigenvalue of the model Hamiltonian. However, the model parameters used throughout this work were determined using calculations performed at $N_{\max}=11$, so the fitted parameters and the corresponding $N_{\max}=11$ masses form a self-consistent finite-basis implementation of the model. Replacing these masses by the extrapolated values while retaining parameters fitted at $N_{\max}=11$ would mix two different levels of numerical approximation. A fully consistent infinite-basis calculation would require extrapolating the states entering the fit, refitting the model parameters using those extrapolated values, and then applying the same procedure to the remaining spectrum.

Moreover, the extrapolation used here is empirical and has been examined only for the ground-state $S$-wave states. We therefore use the $N_{\max}\rightarrow\infty$ values as a diagnostic of basis convergence rather than as the final masses quoted in the spectrum. The difference between the $N_{\max}=11$ results and the extrapolated limits is only about $7$--$10$ MeV for the states considered. Finally, because the Hamiltonian itself is an effective phenomenological model that neglects relativistic effects, coupled-channel dynamics, and higher-order spin-dependent interactions, even the infinite-basis limit should not
be interpreted as an exact physical prediction.

\subsection{Predictive Power}
As it stands, the model is at best a semi-empirical phenomenological picture of heavy quarkonium for which the parameters are fitted to the available data, with no apparent predictive power. In this section, we will show that the model can go beyond mere fitting and can, given a small subset of states, predict the masses of higher states and even single out outliers and non-$q\bar{q}$ candidates.

We fit only a small subset of well-established, low-lying, states, namely the $1^1S_0~(\eta_c)$, $1^3S_1~(J/\psi)$, $1^3P_J~(J=0,1,2)$, and $1^3D_J~(J=1,2,3)$ multiplets for charmonium and the $1^1S_0~(\eta_b)$, $1^3S_1~(\Upsilon)$, $1^3P_J~(J=0,1,2)$ multiplet, and $1^3D_J~(J=1,2)$ for bottomonium. The resulting parameters, shown in Table \ref{tab:reduced_params}, were then held fixed while the masses of the remaining states were computed without any additional adjustments. Despite using a few states in the fit, the model reproduces most higher-lying conventional states to within approximately $1\%$. While the average deviation is higher than the one obtained by fitting the full spectrum ($69.8$ MeV versus $26.8$ MeV for charmonium and $24.3$ MeV versus $14.6$ MeV for bottomonium), the largest discrepancies occur for states such as $X(3872)$, $Z_c(3900)$, $\psi(4230)$, and $Z_c(4430)$, which are generally considered candidates for exotic hadrons with significant non-$c\bar{c}$ components \cite{Swanson:2004pp,Voloshin:2013dpa,Wang:2025clb} and the $Z_b(10610)$ and $Z_b(10650)$, which are interpreted as 'hadronic molecules' in the literature \cite{Li:2012uc}, but we assume to be both $3^{1}P_1$ states. These non-conventional states, which were omitted in the previous discussion, are labeled by asterisks in the tables. The results, reported in Table \ref{tab:predictive_charmonium} and Table \ref{tab:predictive_bottomonium}, demonstrate that the model possesses genuine predictive power while naturally identifying states that may lie beyond the conventional quark-model description. 
\begin{table}[h]
    \centering
    \begin{tabular}{c|ccccc}
    \hline\hline
      & ~~$m_q$ (GeV)~~ & ~~$\alpha_s$~~ & ~~$b~\text{(GeV)}^2$~~ & ~~$\sigma$ (GeV)~~ & ~~$\mathcal{C}$ (GeV)~~ \\
      \hline
      ~Charmonium~ & 1.420 & 0.474 & 0.160 & 0.715 & 0.001 \\ 
      ~Bottomonium~ & 4.800  & 0.394 & 0.163 & 1.572 & 0.0 \\
      \hline\hline
    \end{tabular}
    \caption{Values of the parameters obtained by fitting the low-lying charmonium and bottomonium states. }
    \label{tab:reduced_params}
\end{table}
\begin{table}[h]
\centering
\begin{tabular}{ccccc}
$~{n}^{2S+1}L_{J}~$ & ~Name (Multiplet)~ & ~Experimental Mass~ & ~Theoretical Mass~ & ~$|\%~\mathrm{error}|$~\\
\hline\hline
$\mathbf{{1}^1S_0}$ & $\mathbf{\eta_c(1S)}$ & \textbf{2.9839} & \textbf{2.9817} & \textbf{0.07} \\
$\mathbf{{1}^3S_1}$ & $\mathbf{J/\psi(1S)}$ & \textbf{3.0969} & \textbf{3.0809} & \textbf{0.52} \\
\hline
$\mathbf{{1}^3P_0}$ & $\mathbf{\chi_{c0}(1P)}$ & \textbf{3.4147} & \textbf{3.5168} & \textbf{2.99} \\
$\mathbf{{1}^3P_1}$ & $\mathbf{\chi_{c1}(1P)}$ & \textbf{3.5107} & \textbf{3.5135} & \textbf{0.08} \\
${1}^1P_1$ & $h_c(1P)$ & 3.5254 & 3.5044 & 0.59 \\
$\mathbf{{1}^3P_2}$ & $\mathbf{\chi_{c2}(1P)}$ & $\mathbf{3.5562}$ & $\mathbf{3.5070}$ & $\mathbf{1.38}$ \\
\hline
${2}^1S_0$ & $\eta_c(2S)$ & 3.6375 & 3.6622 & 0.68 \\
${2}^3S_1$ & $\psi(2S)$ & 3.6861 & 3.6726 & 0.37 \\
\hline
$\mathbf{{1}^3D_1}$ & $\mathbf{\psi(3770)}$ & \textbf{3.7737} & \textbf{3.8087} & \textbf{0.93} \\
$\mathbf{{1}^3D_2}$ & $\mathbf{\psi_2(3823)}$ & \textbf{3.8237} & \textbf{3.8017} & \textbf{0.57} \\
$\mathbf{{1}^3D_3}$ & $\mathbf{\psi_3(3842)}$ & \textbf{3.8427} & \textbf{3.7912} & \textbf{1.34} \\
\hline
${2}^3P_1*$ & $X(3872)$ & 3.8717 & 3.9580 & 2.23 \\
${2}^1P1*$ & $Z_c(3900)$ & 3.8871 & 3.9528 & 1.69 \\
${2}^3P_0$ & $X(3915)$ & 3.9217 & 3.9609 & 1.00 \\
${2}^3P_2$ & $\chi_{c2}(2P)$ & 3.9223 & 3.9522 & 0.76 \\
\hline
${3}^3S_1$ & $\psi(4040)$ & 4.0390 & 4.0995 & 1.50 \\
${2}^3D_1$ & $\psi(4160)$ & 4.1910 & 4.1946 & 0.08 \\
${3}^3D_1*$ & $\psi(4230)$ & 4.2200 & 4.5333 & 7.42 \\
${3}^3P_1$ & $\chi_{c1}(4274)$ & 4.2740 & 4.3168 & 1.00 \\
${3}^3S_1*$ & $\psi(4360)$ & 4.3680 & 4.5333 & 3.78 \\
${4}^3S_1$ & $\psi(4415)$ & 4.4210 & 4.4586 & 0.85 \\
${3}^1P_1*$ & $Z_c(4430)$ & 4.4780 & 4.3258 & 3.40 \\
${5}^3S_1$ & $\psi(4660)$ & 4.6300 & 4.7800 & 3.24 \\
\hline\hline
\end{tabular}
\caption{Comparison of experimental and theoretical charmonium masses (in GeV) obtained using parameters determined from the low-lying states shown in bold. These states were included in the fit, while all remaining states were predicted without any further adjustment of the model parameters. The average deviation over the full spectrum is approximately 69.8 MeV. The largest discrepancies occur for several states widely considered candidates for exotic hadrons or states with significant non-$c\bar c$ components, denoted with an asterisk.}
\label{tab:predictive_charmonium}
\end{table}
\begin{table}[h]
\centering
\begin{tabular}{ccccc}
$~{n}^{2S+1}L_{J}~$ & ~Name (Multiplet)~ & ~Experimental Mass~ & ~Theoretical Mass~ & ~$|\%~\mathrm{error}|$~\\
\hline\hline
$\mathbf{{1}^1S_0}$ & $\mathbf{\eta_b(1S)}$ & \textbf{9.3987} & \textbf{9.3899} & \textbf{0.09} \\
$\mathbf{{1}^3S_1}$ & $\mathbf{\Upsilon(1S)}$ & \textbf{9.4603} & \textbf{9.4522} & \textbf{0.09} \\
\hline
$\mathbf{{1}^3P_0}$ & $\mathbf{\chi_{b0}(1P)}$ & \textbf{9.8594} & \textbf{9.8837} & \textbf{0.25} \\
$\mathbf{{1}^3P_1}$ & $\mathbf{\chi_{b1}(1P)}$ & \textbf{9.8928} & \textbf{9.8986} & \textbf{0.06} \\
${1}^1P_1$ & $h_b(1P)$ & 9.8993 & 9.9108 & 0.12 \\
$\mathbf{{1}^3P_2}$ & $\mathbf{\chi_{b2}(1P)}$ & \textbf{9.9122} & \textbf{9.9262} & \textbf{0.14} \\
\hline
${2}^1S_0$ & $\eta_b(2S)$ & 9.9990 & 9.9996 & 0.01 \\
${2}^3S_1$ & $\Upsilon(2S)$ & 10.0233 & 10.0034 & 0.20 \\
\hline
$\mathbf{{1}^3D_1}$ & $\mathbf{\Upsilon(1D)}$ & \textbf{10.1611} & \textbf{10.1424} & \textbf{0.18} \\
$\mathbf{{1}^3D_2}$ & $\mathbf{\Upsilon_2(1D)}$ & \textbf{10.1637} & \textbf{10.1476} & \textbf{0.16} \\
\hline
${2}^3P_0$ & $\chi_{b0}(2P)$ & 10.2325 & 10.2253 & 0.07 \\
${2}^3P_1$ & $\chi_{b1}(2P)$ & 10.2555 & 10.2373 & 0.18 \\
${2}^1P_1$ & $h_b(2P)$ & 10.2598 & 10.2483 & 0.11 \\
${2}^3P_2$ & $\chi_{b2}(2P)$ & 10.2686 & 10.2601 & 0.08 \\
\hline
${3}^3S_1$ & $\Upsilon(3S)$ & 10.3552 & 10.3315 & 0.23 \\
${4}^3S_1$ & $\Upsilon(4S)$ & 10.5794 & 10.5952 & 0.15 \\
\hline
${3}^1P_1*$ & $Z_b(10610)$ & 10.6070 & 10.5170 & 0.85 \\
${3}^1P_1*$ & $Z_b(10650)$ & 10.6520 & 10.5170 & 1.27 \\
\hline\hline
\end{tabular}
\caption{Comparison of experimental and theoretical bottomonium masses (in GeV) obtained using parameters determined from the low-lying states shown in bold. These states were included in the fit, while all remaining states were predicted without any further adjustment of the model parameters. The average deviation over the full spectrum is approximately 24.3 MeV. The largest discrepancies occur for the states widely considered candidates for exotic hadrons or states with significant non-$b\bar b$ components, denoted with an asterisk.}
\label{tab:predictive_bottomonium}
\end{table}
\section{Summary and Conclusions}
\label{sec:Summary}
We have developed a simple non-relativistic phenomenological model of heavy quarkonium leading to a radial hydrogen-like Schr\"odinger equation with a Coulomb-plus-linear potential, which is then solved in a finite Gaussian Expansion Method (GEM) basis. The resulting model yields theoretical values for the heavy quarkonium mass spectrum that are in good agreement with experimental observations. The calculations reproduce the overall structure of both the charmonium and bottomonium spectra while illustrating the role of spin-dependent interactions and the differing behavior of states with different orbital angular momentum.

The framework combines several standard techniques of quantum mechanics---matrix diagonalization, variational methods, and perturbation theory---within a physically realistic application. The equations obtained are sufficiently accessible to be tackled by advanced undergraduate students familiar with quantum mechanics, while at the same time providing an opportunity to explore a genuine problem in hadron spectroscopy. In particular, the model illustrates how effective potentials can capture many of the essential features of QCD bound states despite the underlying complexity of the full theory.

While there is some overlap between our model and the earlier work of Belhouari \cite{Belhouari2023}, our framework is more general and detailed. We compute both the charmonium and bottomonium spectra, whereas Ref.~\cite{Belhouari2023} considers only charmonium. In addition, our model includes the spin-orbit interaction, which is neglected in Ref.~\cite{Belhouari2023}. The spin-spin hyperfine interaction removes the degeneracy between states with different total spin $S$, while the spin-orbit interaction lifts the remaining degeneracy among states with different total angular momentum $J$ for fixed orbital angular momentum $L$ and spin $S$.

The model also lends itself naturally to further exploration. For example, students familiar with tensor operators could extend the Hamiltonian by including the tensor interaction and studying its contribution to the fine-structure splittings. Relativistic corrections could also be incorporated and compared with the present non-relativistic treatment. In addition, one could investigate alternative basis choices, such as a true simple harmonic oscillator basis,
\begin{equation}
N_{n\ell}(\beta r)^{\ell}
L_n^{\ell+1/2}(\beta^2 r^2)
e^{-\beta^2 r^2/2},
\end{equation}
and compare the resulting convergence properties with those obtained using the GEM basis employed here.

Lastly, it would be interesting to investigate whether a single global parameter set could simultaneously describe both charmonium and bottomonium. Such an extension would likely require replacing the fixed strong coupling parameter $\alpha_S$ with a running coupling that depends on the characteristic energy scale of the system. By contrast, the confining parameter $b$ already exhibits approximate universality, since its fitted values for charmonium and bottomonium are relatively similar. This suggests that the long-range confining interaction may be less sensitive to the heavy-quark mass scale than the short-range Coulombic interaction.

\section{Authors Declarations}
\textbf{Conflict of Interest}

The authors have no conflict to disclose.

\section{Appendix: Matrix Elements}
\label{sec:appendix}
In this appendix we give explicit expressions for the matrix elements of the various terms in equation \ref{eq:Hmatrixelements}, separated by term for easier readability. The kinetic term, $H_0$ is:
\begin{align}
    H_{0_{mn}}&=\braket{\psi_m|H_0|\psi_n}=\int_0^\infty dr r^2 \psi^*_{m}(r;\beta,\ell)\left(-\frac{1}{m_q}\left[\frac{d^2}{dr^2}+\frac{2}{r}\frac{d}{dr} - \frac{\ell(\ell+1)}{r^2}\right]\right)\psi_{n}(r;\beta,\ell)\\ \nonumber
    &=-\frac{\beta ^2 2^{\ell+\frac{1}{2}} \Gamma \left(\ell+\frac{1}{2}\right) (m n)^{\frac{1}{4} (2 \ell+3)} (m+n)^{-\ell-\frac{5}{2}} \left(m^2 \ell (\ell+1)-m (2 \ell (\ell+3)+3) n+\ell (\ell+1)
   n^2\right)}{m_q \Gamma \left(\ell+\frac{3}{2}\right)},
\end{align}
where $m_q$ is the quark (antiquark) mass, which in the case of heavy quarkonium is the reduced mass of the system, $\Gamma(n)$ is the Gamma function, and $\ell$ is the orbital angular momentum.
The overlap matrix and the $2m_q$ terms have the same expression, since the $\braket{2m_q}=2m_qS_{mn}$. This is given by
\begin{align}
    S_{mn}&=\braket{\psi_m|\psi_n}=\int_0^\infty dr r^2 \psi^*_{m}(r;\beta,\ell)\psi_{n}(r;\beta,\ell)\\ \nonumber
    &=2^{\ell+\frac{3}{2}} (m n)^{\frac{1}{4} (2 \ell+3)} \left(\frac{1}{m+n}\right)^{\ell+\frac{3}{2}}.
\end{align}
The Coulomb-plus-linear potential term $V_{SI}$ is
\begin{align}
    V_{SI_{mn}}&=\braket{\psi_m|V_{SI}|\psi_n}=\int_0^\infty dr r^2 \psi^*_{m}(r;\beta,\ell)\left( -\frac{4}{3}\frac{\alpha_{s}}{r}+br+\mathcal{C}\right)\psi_{n}(r;\beta,\ell)\\ \nonumber
    &=2^\ell (m n)^{\frac{1}{4} (2 \ell+3)} (m+n)^{-\ell} \left(\frac{\Gamma (\ell+1) \left(2 \beta^2 \alpha_S (m+n)-3 b (\ell+1)\right)}{\beta  (m+n)^2 \Gamma
   \left(\ell+\frac{3}{2}\right)}+\frac{2 \sqrt{2} \mathcal{C}}{(m+n)^{3/2}}\right),
\end{align}
Lastly, the hyperfine term is given by
\begin{align}
    V_{hyp_{mn}}&=\braket{\psi_m|V_{hyp}|\psi_n}=\int_0^\infty dr r^2 \psi^*_{m}(r;\beta,\ell)\left( \frac{32\pi a_{s}}{9m^2_q}\left(\frac{\sigma}{\sqrt{\pi}}\right)^3 e^{-\sigma^2 r^2}\right)\psi_{n}(r;\beta,\ell)\\ \nonumber
    &=\frac{\alpha_{S} 2^{\ell+\frac{13}{2}} \sigma ^3 \beta ^{2 \ell+3} (m n)^{\frac{1}{4} (2 \ell+3)} \left(\beta ^2 (m+ n)+2 \sigma ^2\right)^{-\ell-\frac{3}{2}}}{9 \sqrt{\pi } m_q^2}.
\end{align}
These are the expressions that are inserted in the diagonalization routine of Eq. \ref{eq:GenEigen} to obtain the eigenvalues.



\begin{thebibliography}{99}
\bibitem{Griffiths}
D. J. Griffiths, \emph{Introduction to Quantum Mechanics}, 2nd Edition, (Pearson Prentice Hall, Upper Saddle River, New Jersey, 2005).

\bibitem{Townsend}
J. S. Townsend, \emph{A Modern Approach to Quantum Mechanics}, 2nd Edition, (University Science Books, Mill Valley, CA, 2012).

\bibitem{Sakurai}
J. J. Sakurai (and S. F. Tuan editor), \emph{Modern Quantum Mechanics}, (Addison Wiley, Reading, Massachusetts, 1994).

\bibitem{Farina:2020slb}
C.~Farina, H.~Garcia Tecocoatzi, A.~Giachino, E.~Santopinto and E.~S.~Swanson,
Phys. Rev. D \textbf{102} (2020) no.1, 014023
doi:10.1103/PhysRevD.102.014023
[arXiv:2005.10850 [hep-ph]].

\bibitem{Belhouari2023}
A.~Belhouari,
``The mass spectrum of quarkonium using matrix mechanics,''
Am.\ J.\ Phys.\ \textbf{91}, 34--41 (2023).

\bibitem{MeyerSwanson2015}
C.~A. Meyer and E.~S. Swanson,  
``Hybrid mesons,''  
\emph{Progress in Particle and Nuclear Physics}, vol.~82, pp.~21--58, 2015.  
doi: \href{https://doi.org/10.1016/j.ppnp.2015.03.001}{10.1016/j.ppnp.2015.03.001}.  
Available: \url{https://www.sciencedirect.com/science/article/pii/S0146641015000162}

\bibitem{PDG2026}
F.~Takahashi \textit{et al.} (Particle Data Group),
\textit{Review of Particle Physics},
Int.\ J.\ Mod.\ Phys.\ A \textbf{41}, 2630011 (2026). See the \emph{Quark Model} section.

\bibitem{Entem:2025bqt}
D.~R.~Entem, F.~Fern{\'a}ndez, P.~G.~Ortega and J.~Segovia,
[arXiv:2504.07897 [hep-ph]]. This is a very modern, complete, and pedagogical review of the constituent quark model.

\bibitem{GodfreyIsgur1985}
S.~Godfrey and N.~Isgur,  
``Mesons in a relativized quark model with chromodynamics,''  
\emph{Physical Review D}, vol.~32, no.~1, pp.~189--231, 1985.  
doi: \href{https://doi.org/10.1103/PhysRevD.32.189}{10.1103/PhysRevD.32.189}.

\bibitem{IsgurKarl1978}
N.~Isgur and G.~Karl,  
``P-wave baryons in the quark model,''  
\emph{Physical Review D}, vol.~18, no.~11, pp.~4187--4205, 1978.  
doi: \href{https://doi.org/10.1103/PhysRevD.18.4187}{10.1103/PhysRevD.18.4187}.

\bibitem{IsgurKarl1979}
N.~Isgur and G.~Karl,  
``Ground-state baryons in a quark model with hyperfine interactions,''  
\emph{Physical Review D}, vol.~19, no.~10, pp.~2653--2680, 1979.  
doi: \href{https://doi.org/10.1103/PhysRevD.19.2653}{10.1103/PhysRevD.19.2653}.

\bibitem{Bali2001}
G.~S. Bali,
``QCD forces and heavy quark bound states,''
\emph{Physics Reports} \textbf{343}, 1--136 (2001).

\bibitem{JugeKutiMorningstar1999}
K.~J. Juge, J.~Kuti, and C.~Morningstar,
``Ab initio study of hybrid $\bar{b} g b$ mesons,''
\emph{Phys. Rev. Lett.} \textbf{82}, 4400--4403 (1999).

\bibitem{EichtenFeinberg1981}
E.~Eichten and F.~Feinberg,
``Spin-dependent forces in QCD,''
\emph{Phys. Rev. D} \textbf{23}, 2724--2744 (1981).

\bibitem{KangSchnitzer1975}
J.~S.~Kang and H.~J.~Schnitzer,
``Dynamics of light and heavy bound quarks,''
\emph{Phys. Rev. D} \textbf{12}, 841--854 (1975).

\bibitem{Hiyama:2003cu}
E.~Hiyama, Y.~Kino and M.~Kamimura,
Prog. Part. Nucl. Phys. \textbf{51} (2003), 223-307
doi:10.1016/S0146-6410(03)90015-9

\bibitem{Richard:1992uk}
J.~M.~Richard,
Phys. Rept. \textbf{212}, 1-76 (1992)
doi:10.1016/0370-1573(92)90078-E

\bibitem{Arifi:2024mff}
A.~J.~Arifi, L.~Happ, S.~Ohno and M.~Oka,
Phys. Rev. D \textbf{110} (2024) no.1, 014020
doi:10.1103/PhysRevD.110.014020
[arXiv:2401.07933 [hep-ph]].

\bibitem{WangZhao2025}
Q.~Wang and Q.~Zhao,
``A short review of the vector charmonium-like state $\psi(4230)$,''
\emph{Chin.\ Phys.\ Lett.} \textbf{42}, 110201 (2025).
doi:10.1088/0256-307X/42/11/110201.

\bibitem{Ding:2007rg}
G.~J.~Ding, J.~J.~Zhu and M.~L.~Yan,
Phys. Rev. D \textbf{77} (2008), 014033
doi:10.1103/PhysRevD.77.014033
[arXiv:0708.3712 [hep-ph]].

\bibitem{Swanson:2004pp}
E.~S.~Swanson,
Phys. Lett. B \textbf{598} (2004), 197-202
doi:10.1016/j.physletb.2004.07.059
[arXiv:hep-ph/0406080 [hep-ph]].

\bibitem{Voloshin:2013dpa}
M.~B.~Voloshin,
Phys. Rev. D \textbf{87} (2013) no.9, 091501
doi:10.1103/PhysRevD.87.091501
[arXiv:1304.0380 [hep-ph]].

\bibitem{Wang:2025clb}
Q.~Wang and Q.~Zhao,
Chin. Phys. Lett. \textbf{42} (2025) no.11, 110201
doi:10.1088/0256-307X/42/11/110201
[arXiv:2508.05304 [hep-ph]].

\bibitem{Li:2012uc}
X.~Li and M.~B.~Voloshin,
Phys. Rev. D \textbf{86}, 077502 (2012)
doi:10.1103/PhysRevD.86.077502
[arXiv:1207.2425 [hep-ph]].
\end{thebibliography}
\end{document}